\documentclass[fleqn,10pt]{wlscirep}
\usepackage[utf8]{inputenc}
\usepackage[T1]{fontenc}
\usepackage[printonlyused]{acronym} %[nolist]
\usepackage{hyphenat}
\usepackage{xcolor}
\usepackage{pgf}
\usepackage{float}
\usepackage{subcaption}
\usepackage{amssymb}
\usepackage{amsmath}
\usepackage{amsthm}
\usepackage{physics}
\usepackage{amsbsy}
\usepackage{subcaption}

\title{Harmful Conspiracies in Temporal Interaction Networks: Understanding the Dynamics of Digital Wildfires through Phase Transitions}

\author[1,+]{Kaspara Skovli Gåsvær}
\author[3,4,5]{Pedro G. Lind}
\author[2]{Johannes Langguth}
\author[1,6]{\\Morten Hjorth-Jensen}
\author[7] {Michael Kreil}
\author[3,8,+,*]{Daniel Thilo Schroeder}

\affil[1]{Department of Physics and Center for Computing in Science Education, University of Oslo, 0316 Oslo, Norway}
\affil[2]{Simula Research Laboratory, HPC, Oslo, 0164, Norway}
\affil[3]{Department of Computer Science, OsloMet--Oslo Metropolitan University, N-0130 Oslo, Norway}
\affil[4]{Artificial Intelligence Lab, Oslo Metropolitan University, N-0166 Oslo, Norway}
\affil[5]{NordSTAR--Nordic Center for Sustainable and Trustworthy AI Research, Pilestredet 52, N-0166 Oslo, Norway}
\affil[6]{Department of Physics and Astronomy and Facility for Rare Isotope Beams, Michigan State University, East Lansing, 48824 MI, USA}
\affil[7]{Bayerischer Rundfunk, München, 80335, Germany}
\affil[8]{SINTEF, Sustainable Communication Technologies, Oslo, 0373 Norway}
\affil[+]{these authors contributed equally to this work}
\affil[*]{daniel.t.schroeder@sintef.no}


\begin{abstract}
Shortly after the first COVID-19 cases became apparent in December 2020, rumors spread on social media suggesting a connection between the virus and the 5G radiation emanating from the recently deployed telecommunications network. In the course of the following weeks, this idea gained increasing popularity, and various alleged explanations for how such a connection manifests emerged. Ultimately, after being amplified by prominent conspiracy theorists, a series of arson attacks on telecommunication equipment follows, concluding with the kidnapping of telecommunication technicians in Peru. In this paper, we study the spread of content related to a conspiracy theory with harmful consequences, a so-called digital wildfire. In particular, we investigate the 5G and COVID-19 misinformation event on Twitter before, during, and after its peak in April and May 2020. For this purpose, we examine the community dynamics in complex temporal interaction networks underlying Twitter user activity. We assess the evolution of such digital wildfires by appropriately defining the temporal dynamics of communication in communities within social networks. We show that, for this specific misinformation event, the number of interactions of the users participating in a digital wildfire, as well as the size of the engaged communities, both follow a power-law distribution. Moreover, our research elucidates the possibility of quantifying the phases of a digital wildfire, as per established literature. We identify one such phase as a critical transition, marked by a shift from sporadic tweets to a global spread event, highlighting the dramatic scaling of misinformation propagation. Additionally, we argue that the driving forces behind this observed transition are attributed to influential users, who act as catalysts, accelerating the spread of misinformation. Lastly, our data suggest that the characteristics of such events may be predictable, at least in some instances. From this data, we hypothesize that monitoring minor peaks in user interactions, which precede the critical phase culminating in real-world consequences, could serve as an early warning system, aiding in the prediction and potentially the mitigation of digital wildfires.
\end{abstract}

\begin{document}

%%%%%%%%%%%%%%%%%%%%%%%%%%%%%%%%%%%%%%%%%%%%%%%%%%%%%%%%%%%%%%%%%%%%%%%%
% ACRONYMS
%%%%%%%%%%%%%%%%%%%%%%%%%%%%%%%%%%%%%%%%%%%%%%%%%%%%%%%%%%%%%%%%%%%%%%%%
% % \begin{acronym}
\acrodef{dw}[DW]{Digital Wildfire}
\acrodefplural{dw}[DWs]{Digital Wildfires}
\acrodef{osn}[OSN]{online social network}
\acrodefplural{osn}[OSNs]{online social networks}
\acrodef{annd}[ANND]{average nearest neighbour degree}
\acrodefplural{annd}[ANNDs]{average nearest neighbour degrees}
\acrodef{dnn}[DNN]{deep neural network}
\acrodefplural{dnn}[DNNs]{deep neural networks}
% \end{acronym}
% 

\flushbottom

%%%%%%%%%%%%%%%%%%%%%%%%%%%%%%%%%%%%%%%%%%%%%%%%%%%%%%%%%%%%%%%%%%%%%%%%
% TITLE
%%%%%%%%%%%%%%%%%%%%%%%%%%%%%%%%%%%%%%%%%%%%%%%%%%%%%%%%%%%%%%%%%%%%%%%%
\maketitle
\thispagestyle{empty}

% \noindent Please note: Abbreviations should be introduced at the first mention in the main text – no abbreviations lists. Suggested structure of main text (not enforced) is provided below.

%%%%%%%%%%%%%%%%%%%%%%%%%%%%%%%%%%%%%%%%%%%%%%%%%%%%
\section*{Introduction and Background}
%%%%%%%%%%%%%%%%%%%%%%%%%%%%%%%%%%%%%%%%%%%%%%%%%%%%
% Before the internet
Before the advent of the internet, people primarily relied on print media and radio as their main sources of news. During that era, information dissemination was characterized by a clear separation between the source and the consumer. The flow of information was unidirectional and slow-paced, and it was common practice to trust journalists and newspapers as authoritative and reliable sources of information.

The inception of the internet~\cite{internet_birth}on January 1, 1983, marked a turning point in the way we exchange and receive information. Initially, the Internet was primarily populated by users with expertise in science, technology, engineering, and mathematics. At this stage, discussions were restricted to a small audience, and traditional broadcasting media outlets still held sway. However, a significant shift occurred in the late 1990s with the introduction of the first \acp{osn} like Bolt.com~\cite{peattie2007internet} or SixDegrees.com~\cite{boyd2007social}, which opened the gates for diverse individuals to participate in a broader online discourse. With the ensuing absence of a clear separation between information sources and consumers, the issue of the trustworthiness of the sources started to become more pressing.

The reliability of news agencies today can vary widely depending on the country and agency in question. Nevertheless, news agencies are generally held to a higher level of accountability than \acp{osn}. While fact-checking of social media content may occur in rare cases, it typically only happens after posts potentially reach a large audience~\cite{ifcnreport}, and even then, the desired effect often fails to materialize~\cite{clayton2020real}. Despite these efforts, a remaining challenge for fact-checking is the extreme amount of data available online. Although estimates for the amount of produced data vary depending on the source and the definition of 'data', it is estimated that the amount of data produced globally has been growing exponentially in recent years. According to a report by Seagate and IDC from 2020~\cite{data2020put}, the global datasphere - which includes all the data created, captured, and replicated in a year - was projected to reach 175 zettabytes by 2025. Even though \ac{osn} data is only a smaller part of this, with this amount of information, it is impossible to manually fact-check; thus, misinformation often spreads unnoticed, especially on social media. We argue that the fact that (1) anyone, regardless of their qualifications, can post about anything online, (2) the resulting sheer amount of unchecked misinformation, and (3) the lack of accountability imposed on the providers of \acp{osn}, turns the sea of online information into a maze; tricky to navigate even if aware of these challenges, and potentially dangerous if not. 

As a result, misinformation travels at a speed never previously seen, possibly resulting in severe real-world implications~\cite{banaji2019whatsapp, twickel_2012_russia_oil, biswas_2012, fisher2016pizzagate}. According to the World Economic Forum~\cite{howell_2013}, these phenomena are called \acp{dw} and defined as \textbf{the rapid spread of information or rumors, amplified by the power of social media, which can create significant societal and economic damage in a short amount of time. The term  \ac{dw} we use in this paper extends this definition by adding a temporal dimension. Here, \acp{dw} begin with the first social media post that addresses the potentially fast-spreading topic and end after real-world consequences occur.} Langguth et al.~\cite{5Gconnection} have shown that the topic leading to real-world consequences may continue to be discussed even after these consequences occur. Even more, a new \ac{dw} may emerge around the same complex of topics in a different context. However, for this article, we refrain from such an extended definition and stick to defining the lifecycle of a \ac{dw} from the first social media post until after the real-world consequences.

% In order to address the problem of \acp{dw}, it is imperative to devise automated systems that enable the early detection of misinformation with the potential to escalate into a \ac{dw}, and facilitate timely intervention. In the recent past, researchers have extensively worked on automating the detection of misinformation. This work can be distinct into several approaches~\cite{shu2017fake}, including linguistic-based~\cite{chen2015misleading}, visual-based~\cite{gupta2013faking}, user-based~\cite{castillo2011information}, post-based~\cite{ruchansky2017csi, jin2016news}, and network-based detection~\cite{pogorelov2020fakenews, pogorelov2022combining}. While, to the best of our knowledge, there is no scientific work introducing a holistic approach to detect and fight \acp{dw} specifically, it's essential that we first understand the underlying mechanisms and dynamics that drive the spread of \acp{dw} before we can develop effective automated systems for their detection and mitigation. This knowledge is the basis for creating more robust and accurate algorithms for early detection and prevention.

To effectively combat the rise of \acp{dw}, it is vital to establish automated systems capable of early misinformation detection, thereby allowing for prompt interventions. Despite the wealth of research dedicated to the automated detection of misinformation, categorizable into approaches such as linguistic-based~\cite{chen2015misleading}, visual-based~\cite{gupta2013faking}, user-based~\cite{castillo2011information}, post-based~\cite{ruchansky2017csi, jin2016news}, and network-based detection~\cite{pogorelov2020fakenews, pogorelov2022combining}, a comprehensive strategy targeting \acp{dw} specifically remains undeveloped. Thus, before devising effective automated systems, it is imperative to delve deeper into understanding the mechanisms and dynamics fostering the proliferation of \acp{dw}. Acquiring such knowledge lays a crucial foundation for crafting robust and precise algorithms vital for early detection and prevention.

% What do you actually do (more general)
In this article, we aim for a generic approach exploiting not only the content but rather the underlying interactions of a particular \ac{dw}, namely \textit{the 5G and  COVID-19 misinformation event}~\cite{Langguth2022_5g_misinfo}, within the \ac{osn} Twitter to gain knowledge about the properties and dynamics of the spread of \acp{dw} on a societal scale. Specifically, we investigate the evolution of the temporal networks induced by the interactions between Twitter users to uncover the emergence of \ac{dw}.

\begin{figure}[t!]
    \centering
    \includegraphics[width=0.98\textwidth]{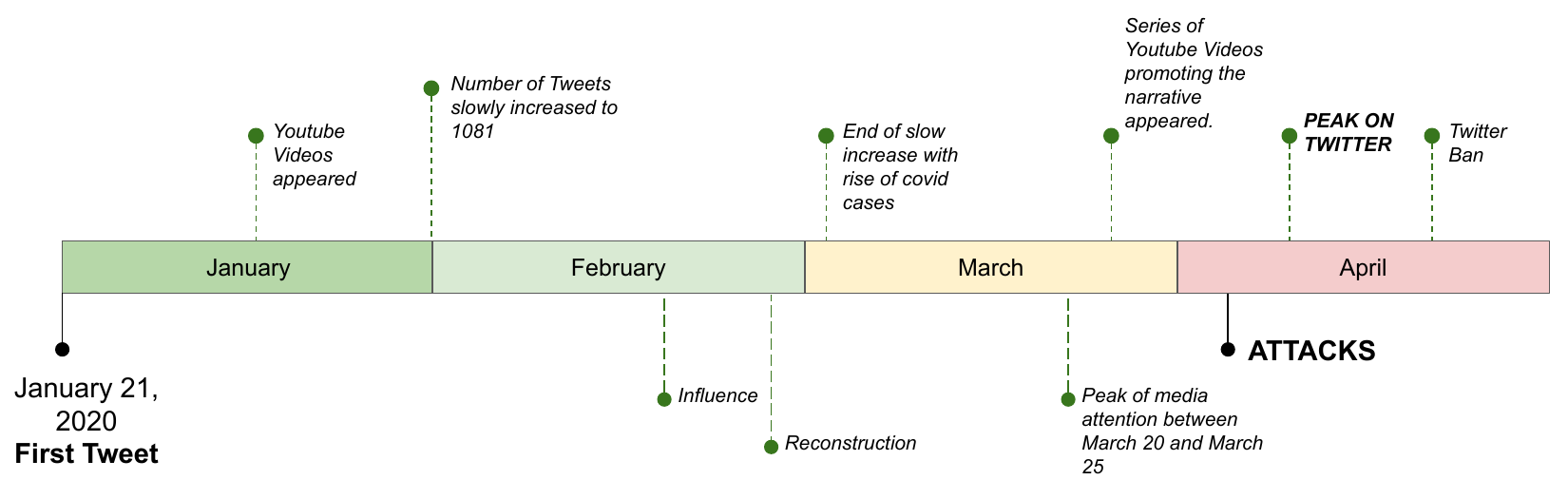}
    % {fig01/fig_timeline_old.pdf}
    \caption{Most significant events that occurred during the course of the COVID-19 and 5G Misinformation event that transpired online in 2020.}
    \label{fig:timeline}
\end{figure}

Previous research has shown that investigating only the diffusion pattern of this kind of misinformation on an individual, per social media post, -basis is not promising at all~\cite{WICOgraph, 5G_mediaevil_2020}. Even more, it seems like we can only understand a \ac{dw} when examining the entirety of information cascades associated with it~\cite{lies_truth_spread_2018}. In this paper, we address the following question:
\textbf{Given the interaction data of an entire \ac{dw} from the online social network Twitter, can we explain its dynamics and temporal evolution on a societal scale by using complex and temporal networks?} The specific temporal network we study originates from interactions between Twitter users connected to the 5G and COVID-19 misinformation event, a series of tweets claiming a link between the COVID-19 virus and 5G technology that lead to a \ac{dw}. This \ac{dw} reached its peak around April $2020$ (see Figure~\ref{fig:timeline}), resulting, among other things, in the destruction of 5G-related telecommunication equipment and the harassment as well as the kidnapping of telecommunication workers.

% What do actually do (a bit more in detail)
% We examine the temporal evolution of the interaction network,~i.e., the spreading of misinformation related to the  5G and COVID-19 event. On Twitter, individual information cascades exist mainly in the form of tweet threads or retweets. Realizing that \acp{dw} consist of a multitude of information cascades, we investigate the evolution of the \ac{dw} considering the entire set of related cascades simultaneously and use community detection~\cite{com_struc_newman, louvain_method, leiden_method} to investigate the dynamics. Moreover, we look into the centrality and activity of the vertices to identify the impact both group- and individual- activity has on the temporal evolution of the network. 
We undertake a examination of the temporal evolution of interaction networks — precisely focusing on the dissemination of misinformation surrounding the 5G and COVID-19 event as it unfolded on Twitter. Individual information cascades on this platform predominantly manifest through threads of tweets and retweets. Understanding that \acp{dw} are essentially conglomerates of numerous such cascades, we venture to scrutinize the evolution of the \ac{dw} through a lens encompassing a significantly comprehensive set of related cascades — a holistic approach facilitated by our comprehensive dataset that, to the best of our knowledge, stands as the most expansive in network-based study of \acp{dw}. Utilizing community detection methodologies~\cite{com_struc_newman, louvain_method, leiden_method}, we delve into discerning the dynamics. Furthermore, by evaluating both the centrality and activity of the vertices, we aim to pinpoint the roles and repercussions of group and individual activities in steering the temporal trajectory of the network.

Our objective is to investigate dynamics to describe and predict the evolution of complex temporal networks related to the spread of \acp{dw}. The insights gained from this study can aid in the early detection and prevention of misinformation events before they escalate into \acp{dw} and end with real-world consequences.

In the following, we list our contributions to the understanding of the 5G and Covid-19 \ac{dw}. First, we demonstrate that the \ac{dw} displays phase transition behavior, highlighting the need to approach this phenomenon from a complex systems and network science perspective. Second, our investigation of community dynamics reveals a synchronization of communities towards the peak of the \ac{dw}, which corresponds to the time when real-world consequences occur. Third, we identify a small group of influential users who are crucial in driving the conversation on a large scale, drawing in a significant number of new users. Finally, our analysis of the largest cluster shows that it is unstable and characterized by oscillations of partly contradictory narratives.

Previous work addresses similar questions and problems. Vosoughi et al.~\cite{lies_truth_spread_2018} investigate the dissemination of true and false news on Twitter, as detailed in their 2018 study. The authors analyze a dataset encompassing approximately 126 thousand news stories tweeted by around three million users, assembled into rumor cascades. Their findings underscore that false news stories propagate significantly faster, reach deeper, and spread more broadly than true stories. Furthermore, false information was found to be retweeted more frequently than true news. Intriguingly, true stories take approximately six times longer to reach a similar audience size of 1,500 users compared to their false counterparts.

Starbird~\cite{starbird2017examining} analyzes Twitter data related to eight mass shooting events from 2013 to 2016, identifying alternative narratives that emerged on the platform. These alternative narratives often contradict mainstream media reports and suggest that the shootings were false flag operations or hoaxes. The study uncovers that alternative media sources play a central role in the production and dissemination of these narratives, acting as key amplifiers of misinformation. The research also highlights the interconnected nature of the alternative media ecosystem, with multiple alternative media sources cross-promoting each other's content and reinforcing the alternative narratives. This interconnectedness contributes to the spread of misinformation and fosters distrust in mainstream media sources.

Del Vicario et al.~\cite{del2016spreading} conducted a study investigating the spreading of misinformation on social media, with a specific focus on Facebook. Their research identifies similar consumption patterns among users who prefer scientific news and conspiracy theories. However, the patterns of information spread, or "cascade dynamics," shows differences. The authors discover that users tend to form "echo chambers," polarized, homogenous clusters where they share content that aligns with their beliefs. Furthermore, the authors introduce a data-driven model that successfully mimicked these dynamics, reinforcing that homogeneity and polarization are key determinants of content spread. 

In their study, Friggeri et al.~\cite{friggeri2014rumor} examine the dynamics of rumor propagation on Facebook. They find that rumors, irrespective of their veracity, spread deeply through social networks, with true rumors generating larger cascades. Their propagation continues even after debunking, indicating that users might overlook or ignore debunking comments. Furthermore, they observe that the popularity of rumors is bursty, with humor sometimes serving as an antidote to rumor propagation. However, despite these findings, the authors acknowledge potential biases in their sample collection and analysis.

%%%%%%%%%%%%%%%%%%%%%%%%%%%%%%%%%%%%%%%%%%%%%%%%%%%%%%%%%%%%%%%%%%%%%%%%%%%%%%%
%%%%%%%%%%%%%%%%%%%%%%%%%%%%%%%%%%%%%%%%%%%%%%%%%%%%%%%%%%%%%%%%%%%%%%%%%%%%%%%

Langguth et al.'s study~\cite{5Gconnection} expanded on the concept of \acp{dw}, focusing on the event we aim to research in this study, the misinformation linking 5G technology with the COVID-19 pandemic. They trace the origin of this rumor and reveal how it grew across social media platforms. The study encounters that even contradictory narratives could strengthen \acp{dw}, and that the role of commercially-influenced videos is often underestimated in Twitter-only analyses. The authors suggest several countermeasures, including focusing on the financial motivations behind the spread of misinformation in general and \acp{dw} in particular and promoting international cooperation in research on \acp{dw}. However, the analyses in this study are more qualitative in nature and do not include a structural analysis of the underlying communication in networks.

%%%%%%%%%%%%%%%%%%%%%%%%%%%%%%%%%%%%%%%%%%%%%%%%%%%%%%%%%%%%%%%%%%%%%%%%%%%%%%%%%%%%
% 5G data as a specific wildfire case
%%%%%%%%%%%%%%%%%%%%%%%%%%%%%%%%%%%%%%%%%%%%%%%%%%%%%%%%%%%%%%%%%%%%%%%%%%%%%%%%%%%%
\section*{COVID-19 and 5G Conspiracy Theories as a Specific Digital Wildfire Case}

As the COVID-19 pandemic swept across the globe in early 2020, a proliferation of tweets emerged linking the virus's origins to 5G wireless technology. Initially confined to a small and insignificant number, the volume of such tweets surged exponentially throughout April 2020, culminating in a series of arson attacks on 5G towers in multiple countries, including the United Kingdom~\cite{kelion_2020}, Nigeria~\cite{staff_2020}, and Canada~\cite{lamoureux_2020}. As mentioned in the previous section, formally, such fast-growing dissemination of online misinformation leading to real-world implications is known as a \acf{dw} and ranked as a top global risk by the World Economic Forum~\cite{howell_2013}. 

In the following, we introduce the chronology of the 5G and COVID-19 misinformation event delineated into three distinct phases, a classification grounded in a qualitative evaluation of the \ac{dw}: pre-real world events, during-real world events, and post-real world events. The demarcation of these phases serves as a framework for unraveling the intricate dynamics at play. Notably, the event persists to this day, warranting continued scrutiny. However, this study is confined to exploring the events until May 2020. For a more detailed overview of the \ac{dw} under investigation, including developments stretching to late 2022, we refer to the adept assessment provided by Langguth et al.~\cite{Langguth2022covid}.

\paragraph{Pre-real world event:} With the first tweet collected in early January, we observed a slow growth in daily tweets, insinuating a connection between COVID-19 and 5G throughout January and February 2020. In addition, we note the gradual uptick in traction for such content on platforms beyond Twitter, including notable activity on YouTube. Pinpointing the exact inception point, however, presents a challenge due to the presence of multiple sub-narratives that are arguably leading to the \ac{dw}. Furthermore, we use Twitter data only, leaving open the possibility that discussions regarding the event were initiated on a platform other than Twitter. However, when investigating the early tweets, we discover an entire spectrum of conspiracy narratives claiming a causality between 5G radiation and the coronavirus. Even though these narratives seem to be as diverse as the individuals spreading them, they share the idea that the 5G technology is dangerous, can hurt people, and thus should not be implemented. For detailed descriptions and tweet samples for subnarratives, we point to the datasets published by Konstantin et al.~\cite{WICOtext} and Schroeder et al.~\cite{WICOgraph}. At this point, we would like to draw your attention to the fact that before the end of January 2020, only 685 tweets and 1,081 retweets containing both keywords referencing COVID-19 and 5G appeared on Twitter. This comparatively small number has led to the decision that we limit ourselves to the period from the first of February for the purposes of this study.

\paragraph{During-real world event:} In March, when the pandemic began to take a foothold in Europe, the spread of tweets also picked up its pace, resulting in four times as many tweets from late March to early April. Consequently, the first series of arson attacks happened in the UK, the Netherlands, and New Zealand during the weekend of April 3, 2020. Multiple more followed in the week after, and later some occurred in Canada as well. By July 2, 2020, there were reports of 273 cases of clashes between people who believed in some version of the conspiracy, as well as 121 reports on arson and other types of destruction~\cite{collins_2020}, including the detainment of 8 telecommunication workers in Peru. 

\paragraph{Post-real world event:} In late April of 2020, Twitter banned material and users promoting attacks on 5G infrastructure, and the spreading of content related to the connection seemed to halt. However, even as late as the first quarter of 2021, suspected cases of arson in Africa and Canada~\cite{vodacom_2021, staff_2021} started to occur.\\

%%%%%%%%%%%%%%%%%%%%%%%%%%%%%%%%%%%%%%%%%%%%%%%%%%%%%%%%%%%%%%%
\begin{figure}[t]
    \centering
    \begin{subfigure}[b]{0.48\textwidth}
         \centering
         \includegraphics[width=\textwidth]{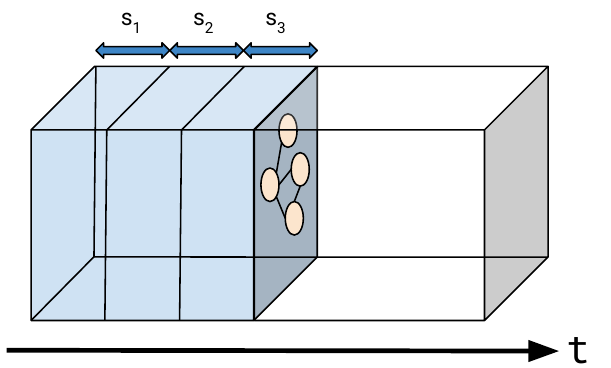}
     \end{subfigure}
     \hfill
     \begin{subfigure}[b]{0.48\textwidth}
         \centering
         \includegraphics[width=\textwidth]{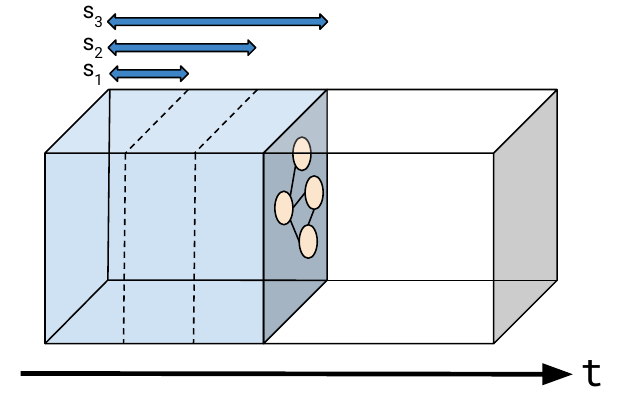}
     \end{subfigure}
     \caption{\protect (Left) Illustration of the concept of temporal slices and
    (Right) accumulative slices.
    Temporal slices do not contain the vertices and edges of previous slices, while accumulative slices contain all vertices and edges from the previous slices.} %\url{https://docs.google.com/drawings/d/162JZ34Q6xmOVQZEEI9yOBWvLI25Lt85pVInN-QVp6KY/edit}}
    \label{fig:slices}
\end{figure}
%%%%%%%%%%%%%%%%%%%%%%%%%%%%%%%%%%%%%%%%%%%%%%%%%%%%%%%%%%%%%%%

\noindent In this paper, we recognize the phase characterized as "during-real world events" as indicative of a phase transition phenomenon, a concept borrowed from the field of statistical physics~\cite{stanleybook,christensenbook}. This classification draws parallels with transitions seen in critical phenomena such as percolation, a well-studied concept in physics. As we delve deeper, it will be evident that the evolution of the \ac{dw} during the COVID-19 and 5G misinformation event exhibits characteristics akin to the onset of a percolation threshold, signifying a critical phase in the information dissemination process.

%%%%%%%%%%%%%%%%%%%%%%%%%%%%%%%%%%%%%%%%%%%%%%%%%%%%%%%%%%%%%%%%%%%%%%%%%%%%%%%%%%%%
% DATABASE
%%%%%%%%%%%%%%%%%%%%%%%%%%%%%%%%%%%%%%%%%%%%%%%%%%%%%%%%%%%%%%%%%%%%%%%%%%%%%%%%%%%%
\section*{Data Collection and Preprocessing of Massive Twitter Datasets}

Since Twitter’s Terms of Service prohibit storing large datasets, we choose a streaming-based approach, first introduced in Schroeder et al.~\cite{schroeder2022connectivity}. We keep only in-stream-anonymized user IDs, the corresponding timestamps, and texts to create the tweet-retweet-user mapping. Moreover, we do neither store nor process any other information. The data collection took place using a custom build framework for Twitter graph analysis~\cite{schroeder2019fact} and a custom scraping strategy~\cite{burchard2020resource}.

Since Twitter's search API, at that time, only returned tweets that were not older than two weeks, it is necessary to collect data preventively while hoping that this collection then, at the time of a \ac{dw}, contains the relevant tweets. In the following, we describe exactly this procedure. 

Between December 2019 and May 2020, we amassed a total of 6,286,886,977 COVID-19-related tweets, retweets, replies, and quotes (referred to as statuses) leveraging the keywords outlined in Appendix~\ref{appendix:keywordsapi} and using Twitter's search API. It is pertinent to note that querying the Twitter search API frequently yields duplicate entries. To ensure the robustness of our dataset, we meticulously identified and removed these duplicates in the initial phase of our data processing. This filtration process resulted in a refined dataset comprising 2,570,581,178 unique statuses, which formed the basis for our subsequent analysis.

Next, we filter for those tweets that mention “ 5G ” and “ 5g ”. We do not remove the whitespaces because doing so produces too many false positives completely unrelated to 5G. Moreover, we include alternative spellings such as “ 5-G ”, although the number of tweets containing these is negligible. All keywords are listed in Appendix~\ref{appendix:keywords5g}. After applying the second filter, $364,325$ Covid-19-related and 5G-related tweets remain. As a concluding procedure, we exclude any statuses not originating from the specific timeframe of February 1, 2020, up to and inclusive of May 11, 2020. \textbf{Hence, our study spans a precise duration of 100 days, beginning on February 1, 2020, and ending on May 11, 2020.}

The enrichment phase commences with the curated dataset derived from the preceding filtering phase. Central to this phase is the intricate process of Twitter thread completion. To elucidate, a Twitter thread is a cohesive series of interconnected tweets stemming from an inaugural tweet and encapsulating all ensuing replies and quoted tweets to foster a consolidated conversation. Such threads offer a structured vantage point, enabling a comprehensive insight into the contextual dynamics enveloping the primary tweet. During this phase, individual threads pertaining to a specified tweet are queried to facilitate the incorporation of statuses surpassing the two-week retrieval constraint imposed by the Twitter search API. Consequently, this method permits the inclusion of tweets devoid of the keywords delineated in Appendix~\ref{appendix:keywordsapi} and~\ref{appendix:keywords5g}. It warrants mention that despite endeavors to augment the dataset by resurrecting more dialogues linked to the \ac{dw}, the potential for thread incompletion persists, attributed to the inherent limitations of the Twitter API, which confines queries to parent statuses within a thread exclusively. Notwithstanding this limitation, we contend that the augmentation process, albeit yielding incomplete threads, substantially enhances the dataset by infusing it with valuable context.

%%%%%%%%%%%%%%%%%%%%%%%%%%%%%%%%%%%%%%%%%
\begin{figure}[t]
    \centering
    \begin{subfigure}[b]{0.33\textwidth}
        \centering
        \includegraphics[width=\textwidth]{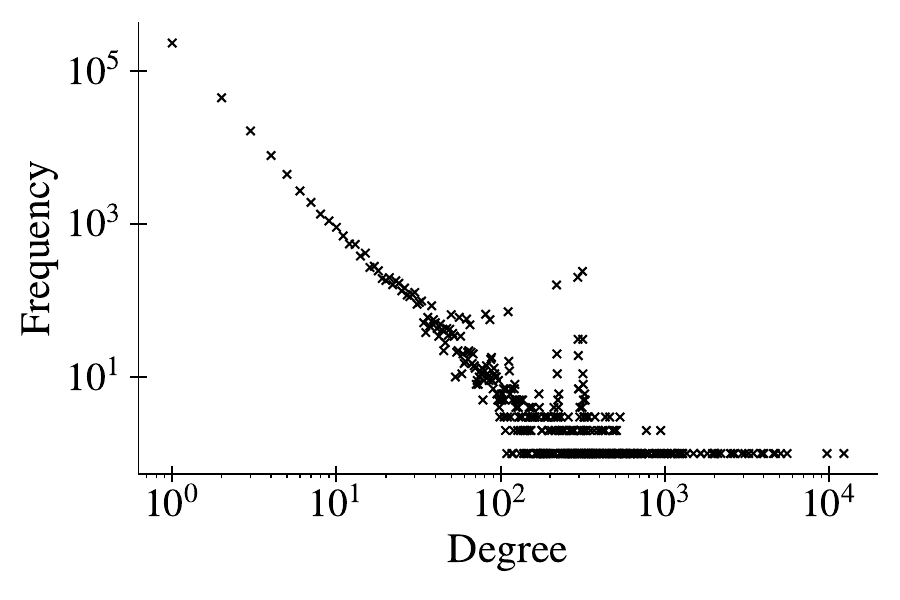}
        % \caption{Global degree distribution.}
    \end{subfigure}
    \begin{subfigure}[b]{0.33\textwidth}
        \centering
        \includegraphics[width=\textwidth]{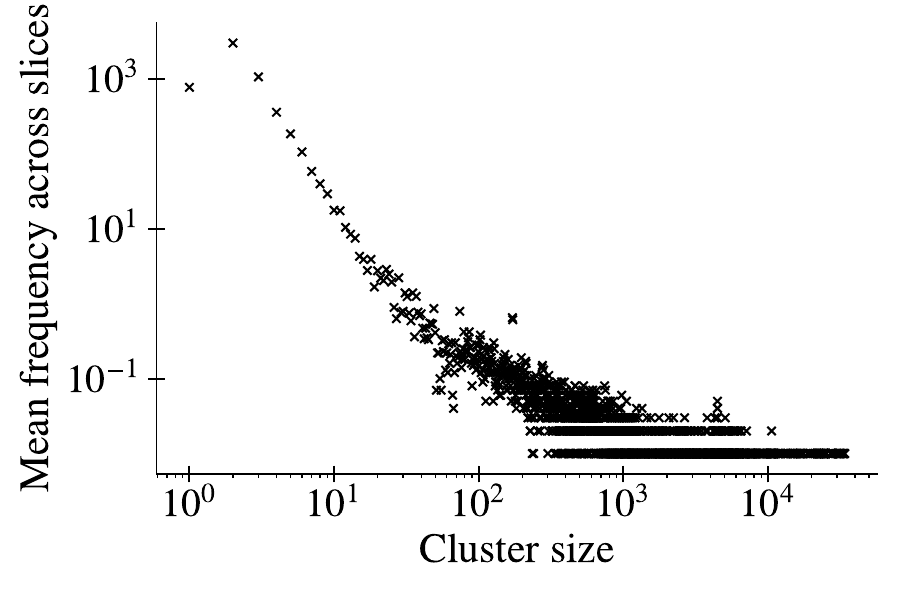}
        % \caption{Accumulative Slices .}
    \end{subfigure}
    \hfill
    \begin{subfigure}[b]{0.33\textwidth}
        \centering
        \includegraphics[width=\textwidth]{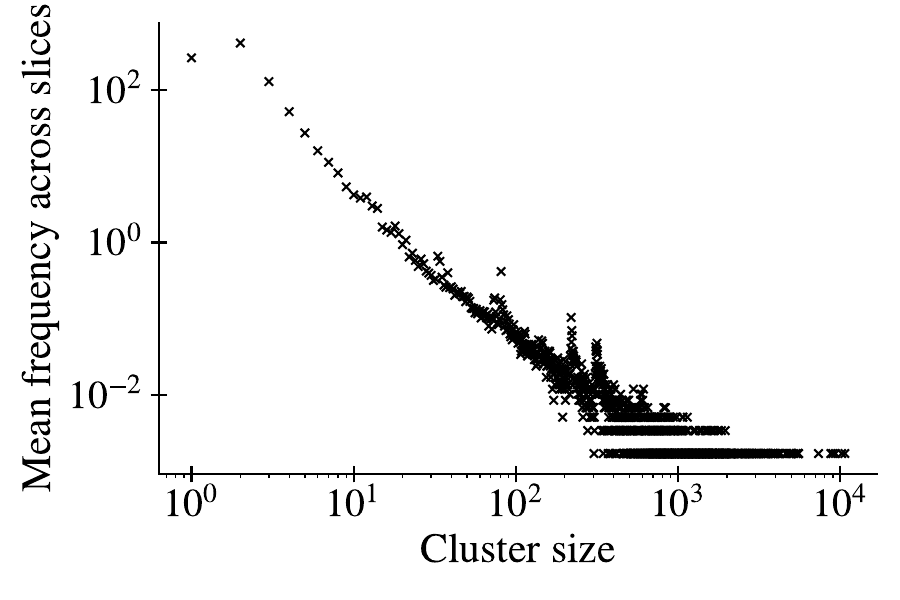}
        % \caption{Temporal Slices $\Delta t = 4h$.}
    \end{subfigure}
    \caption{\protect
     Three-part depiction of Network $G_\downarrow$. The left subfigure presents the global degree distribution, illustrating the range of degree centralities. The central subfigure displays the distribution of cluster sizes across all accumulative slices, with clusters identified via the Leiden algorithm~\cite{traag2019louvain}. The rightmost subfigure explores the cluster size distribution across all temporal slices ($\Delta t = 4h$). Both the middle and rightmost figures are created by counting the occurrence of clusters with size $C$ in every slice before averaging the number of occurrences by the number of slices in the experiment. Notably, both degree and cluster sizes exhibit a power law distribution, mirroring patterns found in other social networks.
    }
     \label{fig:degreedist}
\end{figure}
%%%%%%%%%%%%%%%%%%%%%%%%%%%%%%%%%%%%%%%%%

Following the enrichment phase, we devoted substantial time and resources to delve extensively into the investigation of the resulting dataset, which has fostered a rich bedrock for a plethora of scientific publications. A testament to the rigorous endeavors undertaken in this period is our meticulous labeling effort encompassing more than 9,688 tweets to ascertain whether they were indeed a segment of the 5G COVID-19 misinformation event. This scrupulous initiative bifurcated into two distinct datasets: one housing the tweets, and the other cataloging as many as 3,492 individual tweet-retweet cascades, also labeled to demarcate if a cascade was intertwined with the misinformation event — the datasets are referenced as WICO-Text and WICO-Graph, detailed in publications by Pogorelov et al.\cite{pogorelov2021wico}, and Schroeder et al.\cite{WICOtext}, respectively.

Following the enrichment phase, we engaged in a detailed analysis of the resulting dataset, a process that fostered the foundation for multiple scientific publications. One significant step in this process was the meticulous labeling of over 9,688 tweets to determine their association with the 5G COVID-19 misinformation event. This effort yielded two distinct datasets: one archiving the tweets and another detailing 3,492 individual tweet-retweet cascades, each labeled to indicate association with the misinformation event. These datasets, denominated as WICO-Text and WICO-Graph, are discussed in depth in works by Pogorelov et al.\cite{pogorelov2021wico} and Schroeder et al.\cite{WICOtext}, respectively. To further leverage these datasets, we organised a MediaEval Benchmark Challenge task, wherein both datasets were subdivided into testing and training sets, and distributed to an initial pool of 15 groups. These groups embarked on developing distinct classifiers capable of differentiating between tweets and cascades genuinely associated with the misinformation event. The entire endeavor is documented in detail in~\cite{pogorelov2020fakenews}.

%%%%%%%%%%%%%%%%%%%%%%%%%%%%%%%%%%%%%%%%%%%%%%%%%%%%%%%%%%%%%%%%%%%%%%%%%%%%%%%%%%%%
% METHODS: EXTRACTING TEMPORAL DATA
%%%%%%%%%%%%%%%%%%%%%%%%%%%%%%%%%%%%%%%%%%%%%%%%%%%%%%%%%%%%%%%%%%%%%%%%%%%%%%%%%%%%
\section*{From Temporal Interactions to Interaction Networks}

Given the filtered and enriched dataset, we now extract user interaction by counting contacts between each pair of users. We define $Z_u = \text{set of users}$ and $Z_s = \text{set of statuses}$. A contact between two users is defined as 
\begin{equation}
    \text{any user } j \text{ interacting with any user } i \text{ through user } j \text{ either retweeting, replying, or quoting user } i \text{.}
    \label{eq:contact}
\end{equation}
The set of contacts induces a symmetric adjacency matrix $A$ with $A_{ij} = 1$ to label an existing contact between users $i$ and $j$ and $0$  if such contact does not exist. By keeping track of the number of retweets, replies, and quotes between users, we are able to build the directed and weighted network of interactions. Since in this paper, we focus on assessing the size of connected users, we consider, for simplicity,  the contacts as unweighted and undirected edges, forming the interaction network and its communities. Furthermore, we call such a network temporal interaction network when contacts have timestamps allowing for only considering excerpts of an interaction network within an arbitrary time window. More precisely, we build temporal interaction networks based on the adjacency matrix $A$, defining the \textit{underlying graph} $G_\downarrow$ as the temporal graph containing the entirety of vertices and edges, i.e.
\begin{equation}
    G_\downarrow = (V_\downarrow, E_\downarrow),
    \label{def:UnderlyingGraph}
\end{equation}
where $V_\downarrow = \{v_i\}_{i=0}^N$ is the total set of vertices labeling the users and $E_\downarrow = \{(u,v,t) \vert u,v \in V_\downarrow, t_0 < t \leq t_0+nT\}$, with $T$ the size of the time window during which connections composing the graph occur, $t_0$ is the initial time and $n$ labels the time-window. Notice that both $G_\downarrow$ and $E_\downarrow$ are functions of the time-window
$n$.
%Here, $T$ is the time window of data collection, so that $T = t_{\text{f}} - t_0$. Furthermore, we emphasize that each edge has a unique timestamp $t$, which is the exact datetime of the contact taking place.

%%%%%%%%%%%%%%%%%%%%%%%%%%%%%%%%%%%%%%%%%%%%%%%%%%%%%%%%%%%%%%%%%%%%%%%%%%%%%%%%%%%%
% METHODS: ASSESSING THE EVOLUTION OF INTERACTION NETWORKS
%%%%%%%%%%%%%%%%%%%%%%%%%%%%%%%%%%%%%%%%%%%%%%%%%%%%%%%%%%%%%%%%%%%%%%%%%%%%%%%%%%%%
\section*{Assessing the Evolution of Interaction Networks}

To examine the network dynamics in a temporal way, we slice $G_\downarrow$. Thus, a slice is a subgraph of $G_\downarrow$ and the set including all slices for the observing period is $S = \qty{G(V_\downarrow,E_i), i \in L, E_i \in E_\downarrow}$ where $L$ is the number of slices. Although a set of slices is temporal, each slice is a static "snap-shot" of a time period in the interaction network. In the following, we present two distinct types of slices: accumulative slices and contact slices (see Figure~\ref{fig:slices}). The rationale behind developing multiple slice types is the potential to extract diverse information from each one.

In the context of network analysis, pure temporal slices offer valuable insights into the overall temporal evolution of the system. However, their limitation lies in their inability to trace clusters across multiple slices because of the removal of nonactive vertices in subsequent time periods. Alternatively, accumulative slices provide the advantage of cluster tracking but come with the trade-off of rapidly increasing in size, posing a challenge in handling them effectively.

%%%PL: better first define slices then accumulative slices (which is the summing up of slices... :-))
We define temporal slices (see Figure~\ref{fig:slices} left) and divide the interaction network into slices of sub-graphs based only on the timestamps, and in a non-accumulative manner. Moreover, we remind the reader that edges are contacts, e.g.,~retweets or comments, and thus associated with a timestamp. We divide our set of edges into intervals, e.g., a day, a week, which we call $\Delta t = (t^s, t^e)$ so that we get $L$ slices in total and the entire time interval where we collected our dataset $T = t^e_L -t^s_1$. This results in a temporal graph $\mathcal{G} = (V_\downarrow,E_1,...,E_L)$. Now each slice $s_i \in S$ contains all edges added to the network in the time period $\Delta t_i$ with
\begin{equation}
    S = \qty{G(V_\downarrow,E_i), i \in L}, 
\end{equation}
and $\bigcup_{i=1}^L s_i = G_\downarrow$.

We define accumulative slices (see Figure~\ref{fig:slices} right) as the set of all contacts made in the time interval $\qty[0,t_i], \quad 0 < t_i < t_{i+1} \leq T$, where $T$ is the data acquisition time window. This means that the slice $s_{i+1}$ contains all contacts in $s_i$ plus all other contacts made in the time interval $\qty[t_i,t_{i+1}]$.
\begin{equation}
    s_{i+1} = \qty{u,v \subset V_\downarrow: \exists (u,v,t) \in E_\downarrow \wedge t \in \qty[0,t_{i+1}]}\notag = s_i \cup \qty{u,v \subset V_\downarrow: \exists (u,v,t) \in E_\downarrow \wedge t \in \qty[t_i,t_{i+1}]}.
\end{equation}
Furthermore, we define the distance between two subsequent timestamps, $t_{i+1}-t_i$, as $\Delta t$, which is equal for all $i$. The last slice of the experiment is, by definition, the entire underlying graph, $G_\downarrow$.

%%%%%%%%%%%%%%%%%%%%%%%%%%%%%%%%%%%%%%%%%%%%%%%%%%%%%%%%%%%%%%%%%%%%%%%%%%%%%%%%%%%%
% GRAPH PROPERTIES FOR DESCRIBING THE EVOLUTION OF INTERACTION NETWORKS
%%%%%%%%%%%%%%%%%%%%%%%%%%%%%%%%%%%%%%%%%%%%%%%%%%%%%%%%%%%%%%%%%%%%%%%%%%%%%%%%%%%%
\section*{Degree Centrality as a Proxy for User Activity}
Centrality measures give insight into which users or vertices contribute the most to the flow of information. As we are dealing with an undirected network, we cannot determine whether a vertex is highly active, e.g., comments on other statuses with a high frequency or is made highly active by others, e.g., many other statuses are responses to a status. Thus, we define vertex activity as \textit{the number of contacts a user experiences}. In other words, this is the number of edges connected to a given vertex and thus equal to the degree centrality defined as 
\begin{equation}
    \mathcal{C}_{\text{deg}}(i) = \sum_{i,j\in \mathcal{N}(i)}e_j,
    \label{eq:degree_cent}
\end{equation}
where $\mathcal{N}(i)$ is the neighborhood of the vertex $i$ defined as 
\begin{equation}
    N(v) = \qty{u \subset V_\downarrow, u \neq v : \exists (u,v,t) \in E_\downarrow}.
    \label{eq:N_hood_static}
\end{equation}
Degree centrality can be calculated for each slice $s_{i}$ or the entire network $G\downarrow$. Later, we use this measure to explore the correlation between the overall vertex activity of a group and other properties of the system, such as the size of the largest clusters and the number of overall contacts.

%%%%%%%%%%%%%%%%%%%%%%%%%%%%%%%%%%%%%%%%%%%%%%%%%%%%%%%%%%%%%%%%%%%%%%%%%%%%%%%%%%%%%
% RESULTS: POWER LAWS
%%%%%%%%%%%%%%%%%%%%%%%%%%%%%%%%%%%%%%%%%%%%%%%%%%%%%%%%%%%%%%%%%%%%%%%%%%%%%%%%%%%%%
\section*{Power Law Distribution of User Contacts and Community Sizes}

Many empirical studies have shown that social networks exhibit power-law degree- and community-size distributions. The phenomenon occurs across complex networks ranging from social~\cite{barabasi1999emergence,fortunato2010community} to informational~\cite{faloutsos1999power,adamic2000power} and biological networks~\cite{jeong2000large, li2005towards}. Here, power law distributions arise for various reasons, such as preferential attachment,~i.e., new nodes are more likely to connect to already well-connected nodes; growth,~i.e., networks expand over time; and homophily,~i.e., similar nodes tend to connect with each other.

%%%%%%%%%%%%%%%%%%%%%%%%%%%%%%%%%%%%%%%%%%%%%%%%%%%%%%%%%%%%%%%
%%%%%%%%%%%%%%%%%%%%%%%%%%%%%%%%%%%%%%%%%%%%%%%%%%%%%%%%%%%%%%%%%%%
\begin{figure}[t]
    \centering
        \begin{subfigure}[b]{0.49\textwidth}
         \centering
         \includegraphics[width=\textwidth]{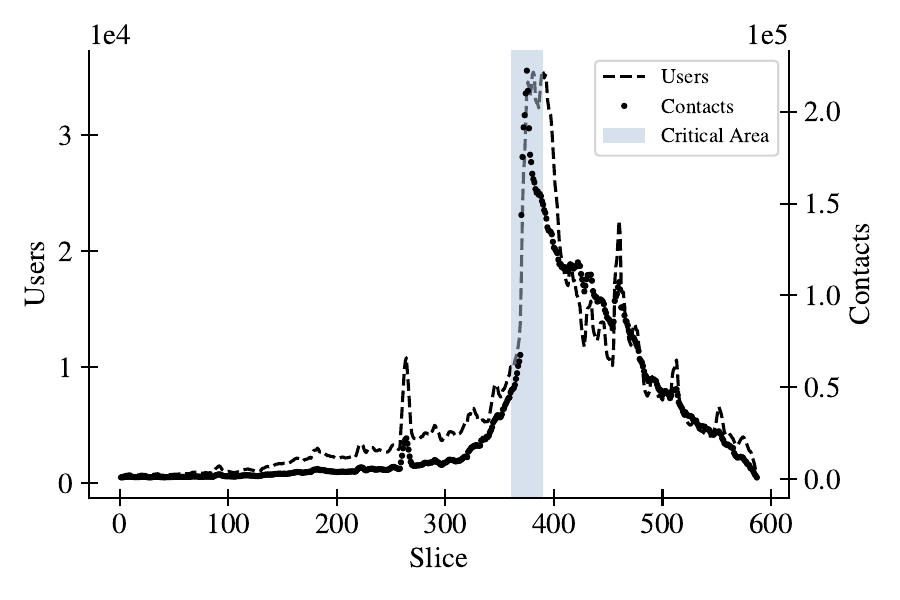}
         % \caption{Temporal slices ($\Delta t = 4h$).}
    \end{subfigure}
    \hfill
    \begin{subfigure}[b]{0.49\textwidth}
         \centering
         \includegraphics[width=\textwidth]{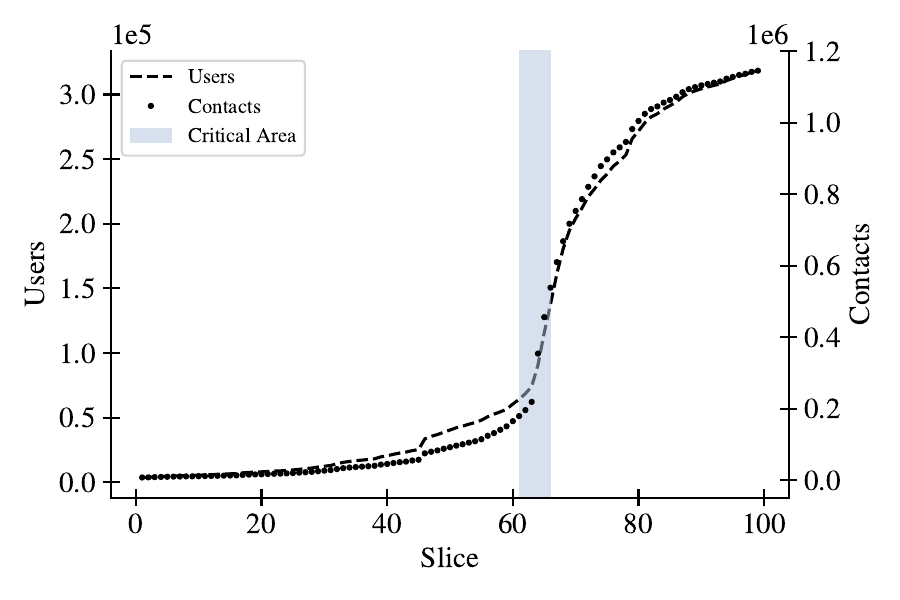}
         % \caption{Accumulative slices ($\Delta t = 24h$).}
    \end{subfigure}
    \caption{
       (Left) Total number of users and established contacts in each time ``slice'' and (Right) the corresponding accumulative slices. Contacts encompass retweets, quotes, and comments, along with the count of distinct users per time segment. The 100-day investigation span, from February 1, 2020, to May 11, 2020, separates into $\Delta t = 24h$ accumulative slices (shown on the right) and $\Delta t = 4h$ temporal slices (shown on the left). Both charts clearly demonstrate a phase transition between slices 360 to 390 and slices 61 to 66, respectively. Additionally, potential predictors emerge between slices 270 and 280 on the right chart. \\
    }
    \label{fig:Nnodes_Ncontacts_exp8}
\end{figure}
%%%%%%%%%%%%%%%%%%%%%%%%%%%%%%%%%%%%%%%%%%%%%%%%%%%%%%%%%%%%%%%%%%%
%%%%%%%%%%%%%%%%%%%%%%%%%%%%%%%%%%%%%%%%%%%%%%%%%%%%%%%%%%%%%%%%%%%

For the internet itself~\cite{barabasi1999emergence} as well as for social networks~\cite{mislove2007measurement}, including Twitter~\cite{schroeder2022connectivity}, it has been shown that both connectivity in general and the number of communication contacts in particular follow power-law distributions. Figure~\ref{fig:degreedist} shows that this also applies to communication within the \ac{dw} under investigation. As with general communication in social networks, there are many users with few contacts to others as well as few users with many contacts to others. Figure~\ref{fig:degreedist} depicts $G_\downarrow$'s global degree distribution, illustrating the range of degree centralities, the community sizes across all accumulative slices, and the community size distribution across all temporal slices. For this, we investigate the community structure underlying our interaction network using the Leiden algorithm~\cite{traag2019louvain} with standard configurations for the resolution parameter.

The power-law degree distributions denote the existence of few hubs, i.e., Twitter users with numerous connections or interactions. Simultaneously, numerous nodes exhibit fewer connections. In the context of \acp{dw}, this distribution suggests that these hubs play a critical role in driving the dynamics of interactions within the \ac{dw}. A minor proportion of users can significantly impact the course of interactions, substantiating their role in the progression of \acp{dw}.

Community size distributions following a power law indicate a composition of multiple small communities alongside a few large ones within the \ac{dw}. While clusters of users exhibit intensive interaction amongst themselves, the network's major interaction activity concentrates within a handful of large communities, for which we later show that conversation within these communities is mostly driven by influential users.

Power law distributions, in degree and community size, bring forth numerous implications for \acp{dw}. The resilience to random node failures associated with power-law degree distributions implies that deactivating or removing a random user might not disrupt the spread of the wildfire. The highly connected hubs maintain the continuum of interaction. Furthermore, we argue that hubs and large communities significantly influence the conversation direction, narrative shape, and information spread within the \ac{dw} while network structure facilitates the rapid and broad diffusion of information, ideas, or behaviors, especially if instigated or promoted by the hubs or large communities.

\begin{figure}
\centering
%\subfloat[]{
\includegraphics[width=.33\linewidth]{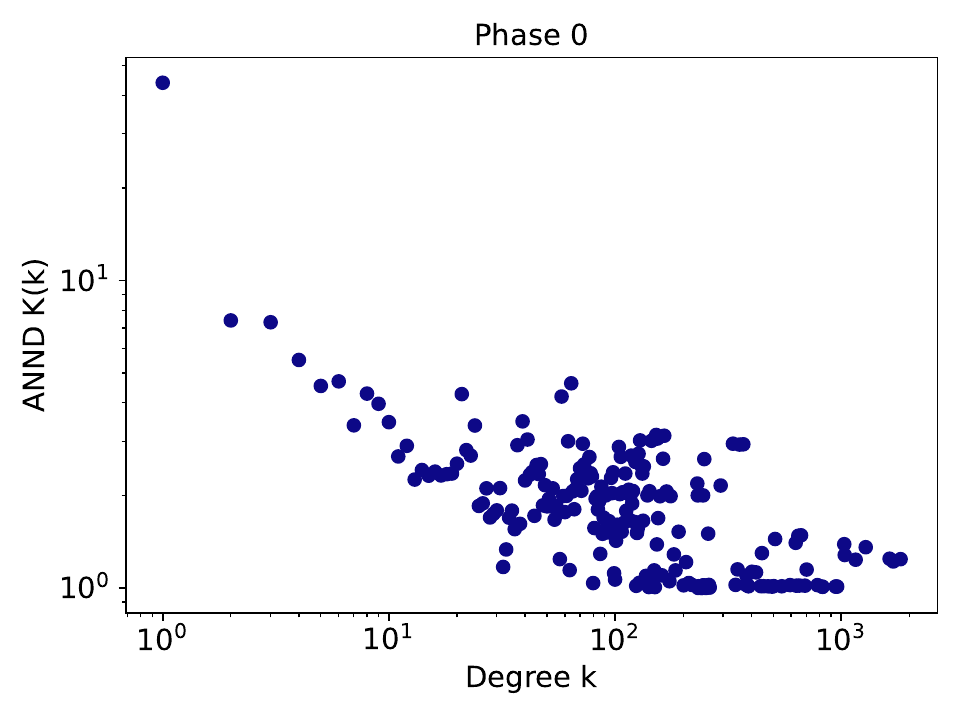}
%\subfloat[]{
\includegraphics[width=.33\linewidth]{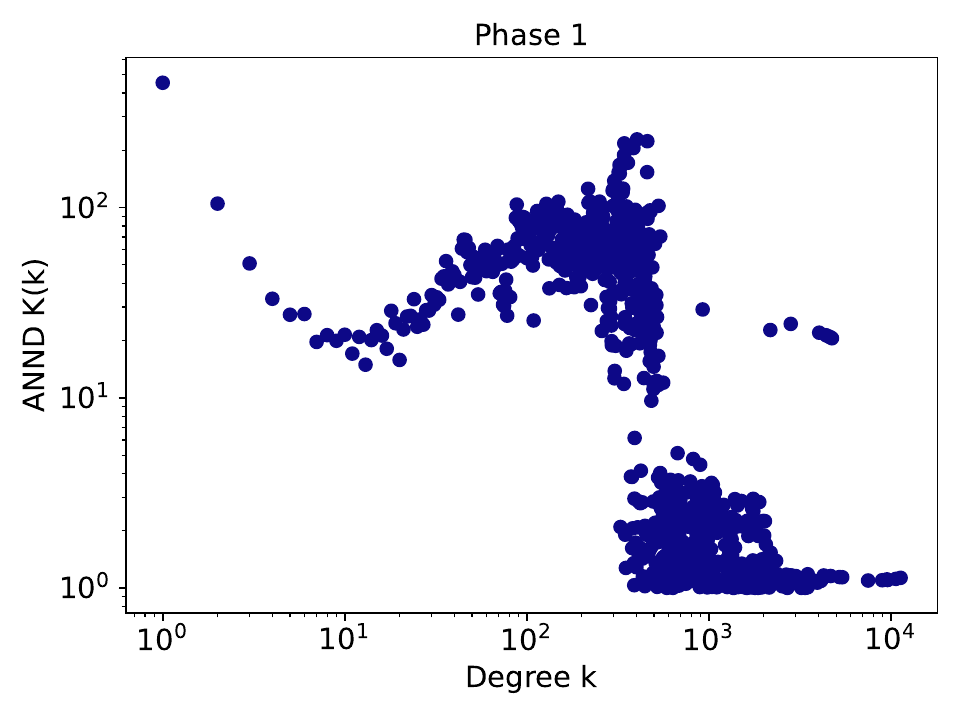}
%\subfloat[]{
\includegraphics[width=.33\linewidth]{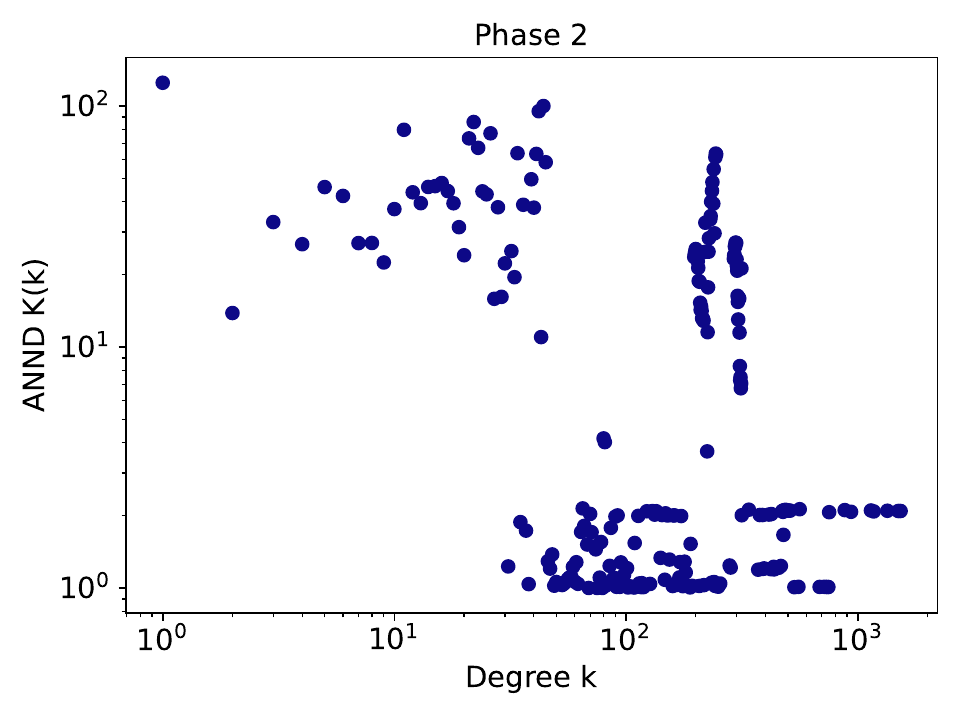}
\caption{The \Acf{annd} function at three different stages:
        (Left) before the phase transition (before slice 360),
        (Middle) during the phase transition (between slices 360 and 390) and
        (Right) after phase transition (after slice 390).}
\label{fig:1}
\end{figure}
%%%%%%%%%%%%%%%%%%%%%%%%%%%%%%%%%%%%%%%%%%%%%%%%%

%%%%%%%%%%%%%%%%%%%%%%%%%%%%%%%%%%%%%%%%%%%%%%%%%%%%%%%%%%%%%%%%%%%%%%%%%%%%%%%%%%%%%
% RESULTS: 3 PHASES + PHASE TRANSITION
%%%%%%%%%%%%%%%%%%%%%%%%%%%%%%%%%%%%%%%%%%%%%%%%%%%%%%%%%%%%%%%%%%%%%%%%%%%%%%%%%%%%%
\section*{Defining the Life Cycle via Phase Transition}

A phase transition is a well-established concept in physics \cite{stanleybook,christensenbook}, with examples such as the transition from a solid to liquid state (melting) or from liquid to gas (evaporation) or the transition from a set of small disconnected groups of individual to a large set of individuals, spanning throughout the entire society. This latter phase transition is usually called transition to percolation~\cite{christensenbook}, and occurs often within the realm of complex systems and network science, usually associated with the emergence of new structures, functionalities, or patterns. Classic examples of phase transitions in complex systems are the emergence of a giant connected component in a random network~\cite{albert2002statistical}, or the abrupt transition from free-flowing traffic to a traffic jam, a scenario often referred to as a 'phantom traffic jam'~\cite{sugiyama2008traffic}. Moreover, phase transitions also serve as effective metaphors for sudden changes in collective human behaviors, particularly in the digital realm~\cite{moussaid2013social}. A specific idea or movement might transition from being recognized by a limited number of individuals to achieving widespread recognition or even reaching viral status.

Qualitatively, a phase transition can be investigated by identifying a parameter whose changes can drive the systems from one phase to another and by keeping track of some observable which characterizes the phase. In the case of liquid-to-gas transition, the parameter is, of course, temperature, and the observable is the density of water. In the case of transition to percolation, the parameter is a sort of probability that pairs of individuals have in establishing one contact, and the observable is the size of the largest group of individuals (cluster). In this section, we argue that the concept of phase transitions is integral to our understanding of~\ac{dw}. In particular, we use it to describe the critical moment when the propagation of misinformation abruptly accelerates, shifting from a slow and steady pace to a rapid, wildfire-like spread. At the same time, the phase transition quantitatively derives the three phases of the \ac{dw} under investigation; see our discussions of Figure~\ref{fig:Nnodes_Ncontacts_exp8}.

The 5G COVID-19 \ac{dw} we analyze in this paper exhibits phase transition characteristics in its spread and contact pattern, suggesting that this phenomenon might apply to other \acp{dw} as well. Figure~\ref{fig:Nnodes_Ncontacts_exp8} on the left side presents temporal slices, specifically time slices containing only temporary interaction networks for the period from February 1, 2020, to May 11, 2020. These slices capture both the number of contacts as described in Equation~\ref{eq:contact} and the visualization of different user counts. Further, a slice (on the left) encompasses the interaction network composed of tweets, retweets, comments, and quotes in time intervals of four hours each. The number of contacts and different users reveals phase transition characteristics between slices 360 and 390, corresponding to the period from April 1 to April 6, 2020. This period precisely precedes and follows the first arson attacks. These observations make it plausible to argue that the categorization of \acp{dw} into three phases, namely before, during, and after real-world consequences, as qualitatively proposed by Langguth et al.~\cite{5Gconnection}, can also be assessed quantitatively.

Figure~\ref{fig:1} shows further evidence that the three stages of a phase transition in time occur during the 5G COVID-19 \ac{dw}. While overall, all three phases appear dominantly disassortative, the average nearest neighbor degree, as well as its variance, significantly vary among these phases.
% , namely \DS{$\langle k_{nn}\rangle = ????$ and $\sigma_{k_{nn}}=????$ before the transition (left), $\langle k_{nn}\rangle = ????$ and $\sigma_{k_{nn}}=????$ at the transition (middle) and $\langle k_{nn}\rangle = ????$ and $\sigma_{k_{nn}}=????$ after the transition (right)}. 
Moreover, Figure~\ref{fig:1} shows the average nearest neighbor degree dependence on the node degree in a qualitatively different way. Before and after the transition, there is a clear degree-range separation: for $k \lesssim 400$ the nearest neighbor degree is typically larger than for $k \gtrsim 400$. At the transition, the nearest neighbor degree seems to distribute more closely to a power-law.
%Moreover, contacts occur in the interval from slice 100 to just after the transition at slice 400 compared to the intervals from 0 to 100 and from 400 to 600. The ratio further increases during the transition, as clearly visible on the right in Figure~\ref{fig:Nnodes_Ncontacts_exp8}. This observation might reflect that fewer users, possibly the same ones, communicate extensively and contribute significantly to the real-world consequences of the \ac{dw}.
This variance could represent an exciting area for future research and a potential unique feature of \acp{dw}. 

%%%%%%%%%%%%%%%%%%%%%%%%%%%%%%%%%%%%%%%%%%%%%%%%%%%%%%%%%%%%%%%%%%%%
\begin{figure}[t!]
    \centering
    \begin{subfigure}[b]{0.48\textwidth}
        \centering
        \includegraphics[width=\textwidth]{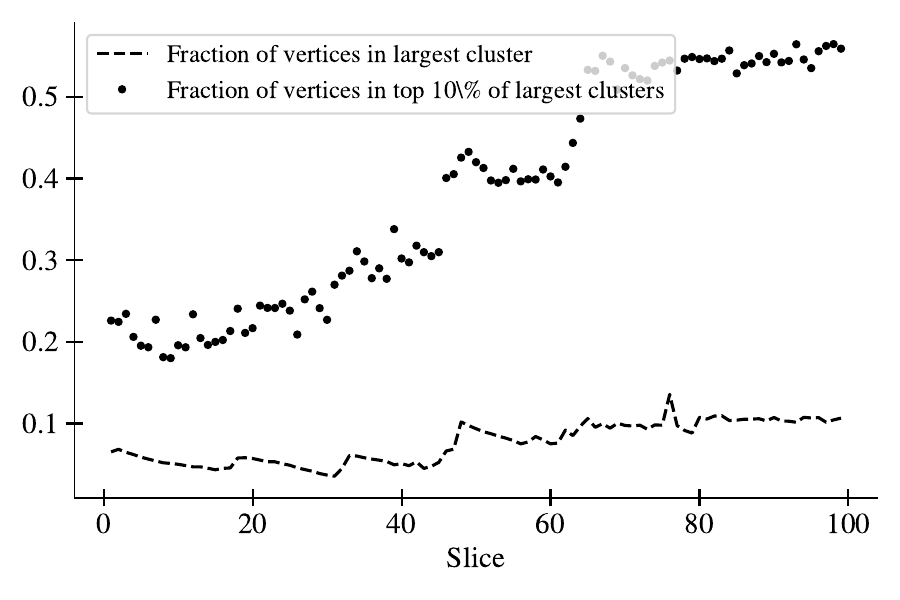}
    \end{subfigure}
    \hfill
    \begin{subfigure}[b]{0.48\textwidth}
        \centering
        \includegraphics[width=\textwidth]{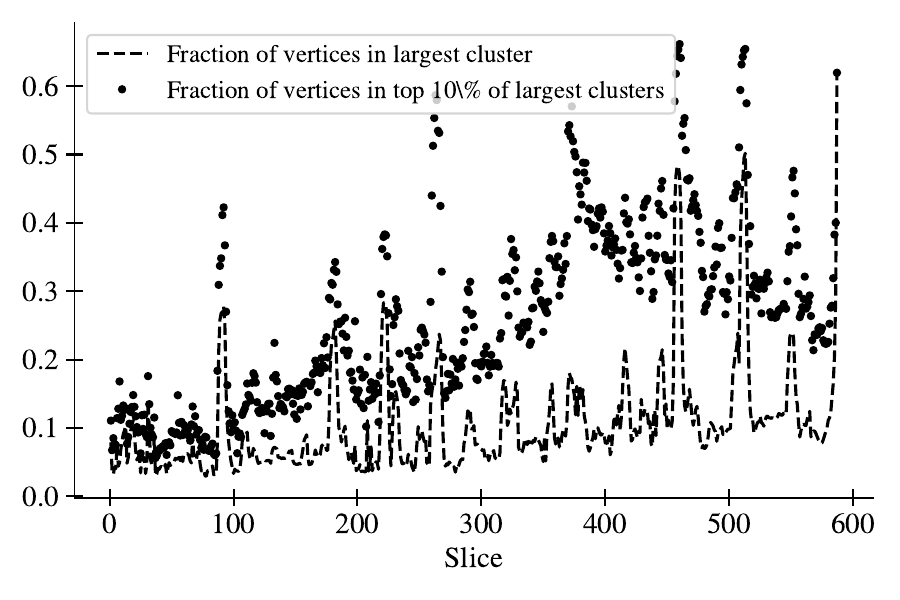}
    \end{subfigure}
    \caption{\protect
     Evolution of user clustering in the analyzed \ac{dw} from February 1, 2020, to May 10, 2020, represented in time slices on the x-axis. The left subplot illustrates the situation with accumulative slices and a $\Delta t = 24h$. It shows a clear and linear increase in the relative size of the largest cluster. This implies a growing participation in the 5G-COVID conspiracy narrative, with more users becoming part of the largest cluster over time. However, the fraction of users who are part of the top 10\% of all clusters does not exhibit such strong growth. The subplot on the right, displaying temporal slices with a $\Delta t = 4h$, shows a similar trend. It is noteworthy that the identity of the largest cluster changes over time, i.e., the largest cluster at a given time does not necessarily contain the same group of users as the largest cluster at a later time.}
    \label{fig:nodes_in_largest_clusters_exp6}
\end{figure}
%%%%%%%%%%%%%%%%%%%%%%%%%%%%%%%%%%%%%%%%%%%%%%%%%%%%%%%%%%%%%%%%%%%%%%%%%%%

%%%%%%%%%%%%%%%%%%%%%%%%%%%%%%%%%%%%%%%%%%%%%%%%%%%%%%%%%%%%%%%%%%%%%%%%%%%%%%%%%%%%%
% RESULTS: Black Hole Transition
%%%%%%%%%%%%%%%%%%%%%%%%%%%%%%%%%%%%%%%%%%%%%%%%%%%%%%%%%%%%%%%%%%%%%%%%%%%%%%%%%%%%%
\section*{Confluence of Coalescing Narratives}

In the study of community dynamics over time, it is crucial to recognize that a community, in this sense, represents more than just a group of users interconnected through high volumes of interaction. In our network, an edge represents an undirected contact, indicating a larger discourse occurring within a particular timeframe. We refer to this discourse as a narrative. This terminology draws from the findings of Langguth et al.~\cite{5Gconnection}, who suggest that \acp{dw}, particularly the 5G COVID-19 \ac{dw}, initially combine followers from various conspiracy narratives. There wasn't one specific source for the \ac{dw}. Instead, various conspiracy theorist groups had already been discussing anti-vaccination and anti-radiation theories prior to the COVID-19 pandemic, which then provided an opportunity for these narratives to merge.

Figure~\ref{fig:nodes_in_largest_clusters_exp6} portrays the evolution of user communities in the analyzed \ac{dw} from February 1, 2020, to May 10, 2020. This is represented as time slices on the $x$-axis. The left subplot, indicating accumulative slices and a $\Delta t = 24h$, shows a clear and linear increase in the relative size of the largest community. However, the proportion of users belonging to the top 10\% of all communities does not increase as significantly. The subplot on the right, which represents temporal slices with a $\Delta t = 4h$, shows a similar trend. Importantly, the composition of the largest community changes over time, meaning the largest community at a given moment may not include the same users as the largest community at a later moment.

When comparing the growth in relative size of the largest community with that of the top 10\% of all communities, we observe that the latter grows at a slower rate. This suggests increasing participation in the 5G-COVID conspiracy narrative, with a growing number of users becoming part of the largest community over time. Therefore, a trend emerges toward the merging of different narratives.

\begin{figure}[t]
    \centering
    \begin{subfigure}[b]{0.48\textwidth}
        \centering
        \includegraphics[width=\textwidth]{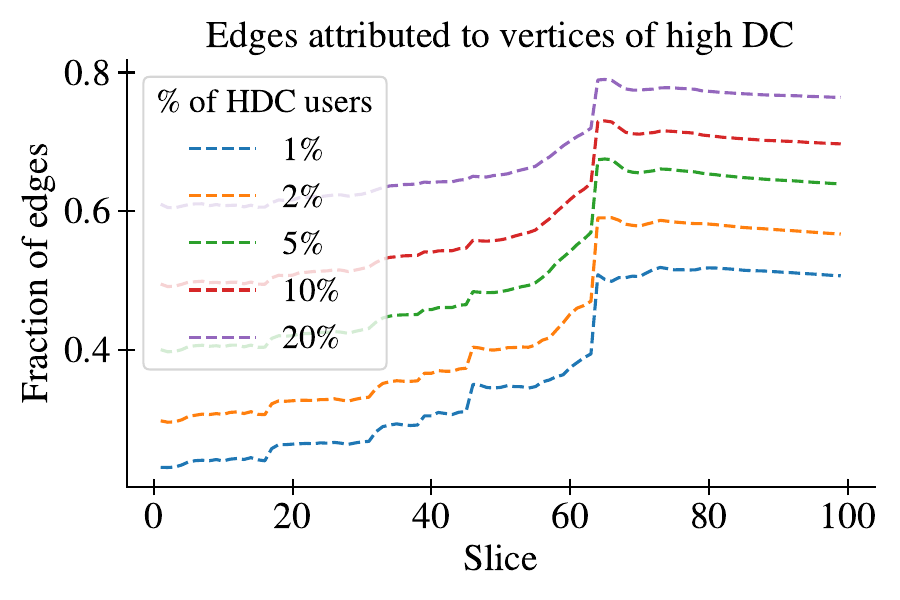}
    \end{subfigure}
    \hfill
    \begin{subfigure}[b]{0.48\textwidth}
        \centering
        \includegraphics[width=\textwidth]{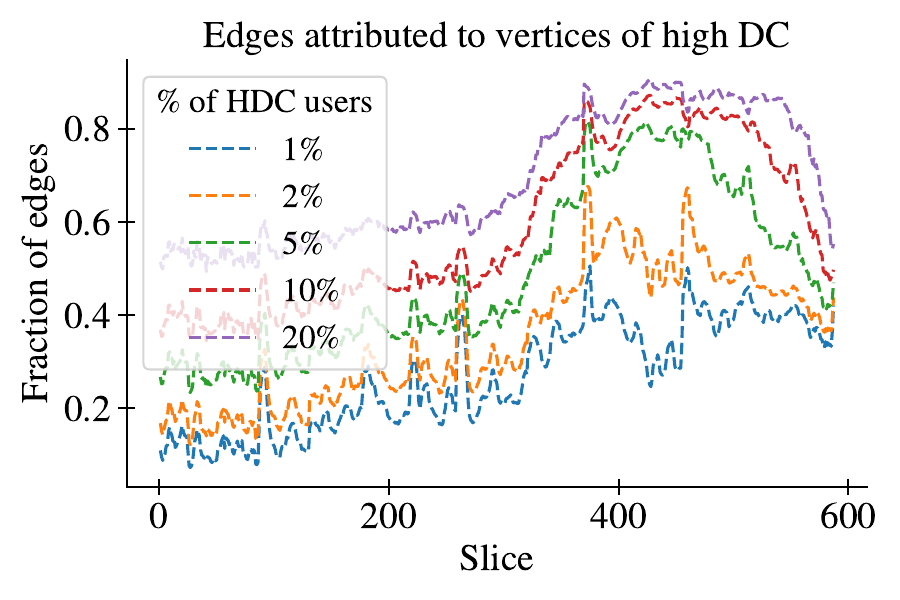}
    \end{subfigure}
    \caption{Correlation between total contacts and the fraction attributable to the most active users, gauged by degree centrality. The left subplot reveals the percentage of vertices in relation to the total number in an accumulative slice with a $\Delta t = 24h$ interval, marking the transition around slice 63. The changing ratio of connections during this transition suggests that highly active users interact with a larger set of distinct users compared to those with lower activity levels. Furthermore, the figure indicates a potential predictor around slice 45, which appears to be primarily noticeable to the active users. The right hand plot presents the same relationship for the temporal slices with $\Delta t = 4h$. Here, we observe similar patterns as on the left. Additionally, the gap becomes apparent between the most active 2\% of users and the less active 5\%, 10\%, or 20\% between slices 350 and 470, underlining the influential role of a small proportion of highly active users in shaping the network dynamics.}
    \label{fig:NAC_vs_num_contacts_exp8}
\end{figure}

%%%%%%%%%%%%%%%%%%%%%%%%%%%%%%%%%%%%%%%%%%%%%%%%%%%%%%%%%%%%%%%%%%%%%%%%%%%%%%%%%%%%%
% RESULTS: Influential Users
%%%%%%%%%%%%%%%%%%%%%%%%%%%%%%%%%%%%%%%%%%%%%%%%%%%%%%%%%%%%%%%%%%%%%%%%%%%%%%%%%%%%%
\section*{Influential Users are Steering the Course of Large-Scale Conversations}

A critical aspect of the dynamic evolution of \acp{dw} is the role of influential users. These users, often characterized by a high degree centrality, have the potential to significantly shape the course of large-scale conversations. Given the network structure of social media platforms like Twitter, a message from an influential user can quickly reach a vast audience, potentially altering the trajectory of an ongoing narrative or sparking a new one.

Figure~\ref{fig:NAC_vs_num_contacts_exp8} illustrates the correlation between total contacts and the fraction attributable to the most active users, gauged by degree centrality. The left subplot reveals the percentage of vertices in relation to the total number in an accumulative slice with a $\Delta t = 24h$ interval, marking the transition around slice 63. The changing ratio of connections during this transition suggests that highly active users interact with a larger set of distinct users compared to those with lower activity levels. Furthermore, the figure indicates a potential predictor around slice 45, which appears to be primarily noticeable to the active users.

The right-hand plot presents the same relationship for the temporal slices with $\Delta t = 4h$. Here, we observe similar patterns as on the left. Additionally, the gap becomes apparent between the most active 2\% of users and the less active 5\%, 10\%, or 20\% between slices 350 and 470, underlining the influential role of a small proportion of highly active users in shaping the network dynamics. Our analysis of the 5G-COVID-19 \ac{dw} reveals that these influential users contribute significantly to the formation and growth of the largest community, as well as to the overall narrative evolution within the \ac{dw}. Notably, we find that users with a high degree centrality have a disproportionate impact on the narrative dynamics, reaching more users compared to those with lower activity levels. Furthermore, our data shows that influential users can act as connectors between different narratives, further fueling the growth of the \ac{dw}. This finding is consistent with the idea that the confluence of coalescing narratives is a fundamental characteristic of \acp{dw}.

In the context of mitigating the impact of harmful \acp{dw}, our findings underscore the importance of monitoring the activity of influential users. Effective strategies could include promoting accurate information through these users or mitigating their influence if they are spreading harmful narratives. Our analysis, however, also highlights the complexity of this task. As the identity of the largest community is not constant over time, so too are the influential users within it. This fluidity necessitates dynamic strategies for monitoring and intervention, tuned to the temporal evolution of the \ac{dw}.

%%%%%%%%%%%%%%%%%%%%%%%%%%%%%%%%%%%%%%%%%%%%%%%%%%%%%%%%%%%%%%%%%%%%%%%%%%%%%%%%%%%%%
% RESULTS: POTENTIAL PREDICTORS
%%%%%%%%%%%%%%%%%%%%%%%%%%%%%%%%%%%%%%%%%%%%%%%%%%%%%%%%%%%%%%%%%%%%%%%%%%%%%%%%%%%%%
\section*{Examination of Pre-transition Phenomena and Exploration of Potential Predictors}

In the realm of \acp{dw}, understanding the dynamics prior to a transition phase can yield critical insights into the potential predictors of such large-scale shifts. Our research illuminates several pre-transition phenomena that suggest impending changes in the narrative or the community structure. We observe a small spike visible around slice 260 in Figure~\ref{fig:Nnodes_Ncontacts_exp8} (left), which portrays the time or time slice on the x-axis and the number of users and user contacts on the y-axis. This peak, preceding the actual transition, could indicate that a topic is gaining potential to generate wider reach, serving as a visible spark that signals an upcoming phase shift.

% \DS{TODO: slice 45 links predictor nur evident bei besonders active benutzern}

Before the transition, we also observe modifications in user behavior patterns. Specifically, the most active users begin to expand their range of contacts. These users, who typically have a high degree centrality, start interacting with a more extensive set of unique users compared to less active ones. This observation reinforces the central role of influential users in guiding the \ac{dw} into a new phase, a notion we discussed in the previous section.

We also see changes in the structural properties of the network before the transition. For instance, the ratio of connections, i.e., the number of communications with different actors, shifts during the transition. This suggests the possibility of using network structure changes as predictive indicators of a \ac{dw} phase transition. These observations suggest that, with further study and refinement, we might be able to establish a predictive model for \ac{dw} transitions. Such a model would not only enhance our understanding of \ac{dw} dynamics but could also provide actionable insights to contain or guide these narratives. Consequently, our future research will focus on refining these potential predictors and testing their predictive power across different contexts.

%%%%%%%%%%%%%%%%%%%%%%%%%%%%%%%%%%%%%%%%%%%%%%%%%%%%%%%%%%%%%%%%%%%%%%%%%%%%%%%%%%%%%
% RESULTS: ROBUST
%%%%%%%%%%%%%%%%%%%%%%%%%%%%%%%%%%%%%%%%%%%%%%%%%%%%%%%%%%%%%%%%%%%%%%%%%%%%%%%%%%%%%
% \section*{TODO: Show which properties of the social network are robust against the during real worl event stage (e.g. size of the largest cluster) and which properties are sensitive to it (e.g. activity)}

% \DS{@PEDRO: I need help here.}

%%%%%%%%%%%%%%%%%%%%%%%%%%%%%%%%%%%%%%%%%%%%%%%%%%%%%%%%%%%%%%%%%%%%%%%%%%%%%%%%%%%%%
% DISCUSSION AND CONCLUSION
%%%%%%%%%%%%%%%%%%%%%%%%%%%%%%%%%%%%%%%%%%%%%%%%%%%%%%%%%%%%%%%%%%%%%%%%%%%%%%%%%%%%%
\section*{Discussion and Conclusion}

In this paper, we address the dynamics and temporal evolution of a specific \ac{dw}, namely the COVID-19 and 5G misinformation event from early 2020. By processing large amounts of Twitter data, we trace the course of this \ac{dw} beyond the investigation of individual information spreading cascades. Rather, we examine the entirety of the underlying communication in interaction networks and thus provide a more holistic view of the dynamics that underlie a \ac{dw}. We show that both the average count of user contacts and the average community sizes over the total duration of the phenomena adhere to a power law distribution. Thus, underline the intrinsic structure and similarity to other communication networks.

Based on the study of these dynamics, we propose a framework that not only allows for accessing the three stages of a \ac{dw} quantitatively through the concept of phase transitions but also allows for the application of well-established methods to understand \acp{dw} through the lens of physics. Physicists are able to predict phase transitions by identifying drivers. 
Temperature, for example, is the driver for the
transition water undergoes between gas, liquid and solid phases. %when evaporating is temperature. 
%Knowing that the study of a single \ac{dw} does not allow general conclusions, we nevertheless 
In this work, we identify potential candidates for drivers in the phase transition of a \acp{dw} based on the underlying communication and thus paving the way towards a \ac{dw} prediction model. 

Thus, we observe a confluence of coalescing narratives by studying the community dynamics over the course of the \ac{dw}. We ask the reader to recall that a community in a time slice of a temporal interaction network is a group of people, in our case, Twitter users, talking to each other (about a topic) within a certain time window. In future work, we plan not only to extend this research to other \acp{dw}, but also to understand the mechanism underlying this confluence, model it, and possibly comprehend it as a building block of a driver. 

Furthermore, we identify a subset of all users involved in the \ac{dw} that could be a major driving force for the transition to a global event. This result suggests that potential drivers can be found in the underlying characteristics of the communication of this user group. As with coalescing narratives, further studies need to show that this result also manifests in other \acp{dw}.

Our study's findings have considerable real-world implications. Unraveling these dynamics can be a potent tool for entities wishing to manage or control information propagation. For democratic societies, these tools can be invaluable in identifying and mitigating nascent extremist groups or misinformation campaigns. On the other hand, they could potentially be exploited to suppress democratic dissent in totalitarian regimes. Given the high stakes, an effective response to \acp{dw} requires an approach that harmonizes technological tools with education, media literacy, and an informed public. Moreover, it's critical to ensure democratic accountability for those who wield these powerful tools.

To conclude, this study represents an effort to model and understand the dynamics of a \ac{dw} within complex temporal interaction networks, with results shedding new light on the societal-scale understanding of these phenomena. Our ultimate objective is to devise methods that can predict and mitigate the spread of harmful misinformation. We encourage future research to validate and expand upon the identified patterns and hypotheses, thus ensuring steady progress toward this paramount objective.

%%%%%%%%%
%%%%%%%%%
%%%%%%%%%
%TEMPORARY FIX FOR GETTING FIGS BEFORE REFS
\clearpage
%%%%%%%%%
%%%%%%%%%
%%%%%%%%%

%%%%%%%%%%%%%%
\bibliography{references}

%%%%%%%%%%%%%%%%%%%%%%%%%%%%%%%%%%
% ACK
%%%%%%%%%%%%%%%%%%%%%%%%%%%%%%%%%%
\section*{Acknowledgements}
This work was funded by the Norwegian Research Council project ”Enabling Graph Neural Networks at Exascale” (\#303404). The research presented in this paper has benefited from the Experimental Infrastructure for Exploration of Exascale Computing (eX3), which is financially supported by the Research Council of Norway under contract \#270053. This work also received partial funding from the projects Graph-Massivizer (HE 101093202), enRichMyData (HE 101070284), and DataCloud (H2020 101016835). In addition, we are indebted to Tore Heide Larsen from the Simula Research Laboratory for managing the infrastructure.

%%%%%%%%%%%%%%%%%%%%%%%%%%%%%%%%%%
% AUTHORS CONTRIBUTION
%%%%%%%%%%%%%%%%%%%%%%%%%%%%%%%%%%
\section*{Author Contributions}
K.S.G., with guidance from D.T.S., conducted the primary analysis and wrote the corresponding sections of the paper, deeply informed by the research undertaken during K.S.G.'s master's thesis. D.T.S. not only supervised K.S.G.'s master's thesis but also led the preprocessing, data enrichment, and took part in both the data collection and manuscript writing. M.K. was responsible for collecting the initial dataset that served as a pivotal resource for this study, laying a substantial foundation for the work presented in this paper. J.L., M.H.J., and P.G.L. were integral to the formative stages of the project, lending their expertise to carve out the creative direction of the study. They remained actively involved through the development of the manuscript, contributing to the writing and meticulously reviewing the content to ensure a polished, cohesive output. All authors reviewed and gave approval for the final version of the manuscript.

%%%%%%%%%%%%%%%%%%%%%%%%%%%%%%%%%%
% COMPETING INTERESTS
%%%%%%%%%%%%%%%%%%%%%%%%%%%%%%%%%%
\section*{Data Availability}
In accordance with Twitter's Terms of Service and to ensure user privacy, we will not publish or share the underlying Twitter source data collected during our study. Our methodology involved the storage of in-stream-anonymized user IDs, timestamps, and the texts necessary to form the tweet-retweet-user mapping, strictly omitting any other personal information.

However, we have created interaction networks derived from the processed Twitter data, which elucidates the structural and relational aspects that informed our analysis, while being compliant with privacy and ethical guidelines. These interaction networks, which formed the foundation of our research, encapsulate the connectivity and dynamics pertaining to the specific period of study from February 1, 2020, to May 10, 2020 and are available under \url{https://osf.io/vqhet/?view_only=6b67055f78e047349036332f5bab7365}. The corresponding codebase is available under \url{https://github.com/KasparaGaasvaer/MasterThesis}.

%%%%%%%%%%%%%%%%%%%%%%%%%%%%%%%%%%
% COMPETING INTERESTS
%%%%%%%%%%%%%%%%%%%%%%%%%%%%%%%%%%
\section*{Competing Interests}
The authors declare no competing interests.

%%%%%%%%%%%%%%%%%%%%%%%%%%%%%%%%%%
% ADD> INFO
%%%%%%%%%%%%%%%%%%%%%%%%%%%%%%%%%%
\section*{Additional Information}
Correspondence and requests for materials should be addressed to D.T.S.

%%%%%%%%%%%%%%%%%%%%%%%%%%%%%%%%%%
% ADD> INFO
%%%%%%%%%%%%%%%%%%%%%%%%%%%%%%%%%%
\section*{Ethical Approval}
The described data collection, anonymization, storage, and use have been reviewed by the Ethical Committee of the Simula Research Laboratory in Oslo and are in accordance with the relevant guidelines and regulations. We see no ethical concerns with the data use or the publication of results based on the data.

\appendix

\section{Keywords for initial Twitter API search}\label{appendix:keywordsapi}
coronavirus10, corinavirus6, corona, coronaoutbreak, coronavirus, coronavirusde, coronavirusoutbreak, covid, covid19, covid2019, covid\_19, covid-19, wuhancoronavirus, wuhanvirus9, coronavírus, coronavirus7, coronavirus8, coronavirus9, zerocovid, coronar\_allesoeffnen, allesoeffnen, allesöffnen, coronadeutchland, xj621, machtbueroszu, machtdiebueroszu, bueroszu, büroszu, diebüroszu, vaccination, vaccine, epidemic, pandemic, quarantine, quarantined, mutation, wuhan, coronapanik, covidiot.

\section{Keywords for filtering 5G-related tweets}\label{appendix:keywords5g}
5G, 5g, 60Hz, \#5G 

%%%%%%%%%%%%%%%%%%%%%%%%%%%%%%%%%%%%%%%%%%%%%%%%%%%%%%%%%%%%%%%%%%%%%%%%%%%
%%%%%%%%%%%%%%%%%%%%%%%%%%%%%%%%%%%%%%%%%%%%%%%%%%%%%%%%%%%%%%%%%%%%%%%%%%%
\end{document}